# Uniaxial Strain Tuning of Charge Modulation and Singularity in a Kagome Superconductor

Chun Lin,[1,*,†] Armando Consiglio,[2,3] Ola Kenji Forslund,[1,4] Julia Küspert,[1] M. Michael Denner,[1] Hechang Lei,[5,6] Alex Louat,[7] Matthew D. Watson,[7] Timur K. Kim,[7] Cephise Cacho,[7] Dina Carbone,[8] Mats Leandersson,[8] Craig Polley,[8] Thiagarajan Balasubramanian,[8] Domenico Di Sante,[9] Ronny Thomale,[2] Zurab Guguchia,[10] Giorgio Sangiovanni,[2] Titus Neupert,[1] and Johan Chang[1]

[1]*Physik-Institut, Universität Zürich, Winterthurerstrasse 190, Zürich CH-8057, Switzerland*[†]
[2]*Institut für Theoretische Physik und Astrophysik and Würzburg-Dresden Cluster of Excellence ct.qmat, Universität Würzburg, Würzburg 97074, Germany*
[3]*Istituto Officina dei Materiali, Consiglio Nazionale delle Ricerche, Trieste I-34149, Italy*
[4]*Department of Physics and Astronomy, Uppsala University, Box 516, Uppsala SE-75120, Sweden*
[5]*School of Physics and Beijing Key Laboratory of Optoelectronic Functional Materials & MicroNano Devices, Renmin University of China, Beijing 100872, China*
[6]*Key Laboratory of Quantum State Construction and Manipulation (Ministry of Education), Renmin University of China, Beijing, 100872, China*
[7]*Diamond Light Source Ltd, Harwell Science and Innovation Campus, Didcot OX11 0DE, United Kingdom*
[8]*MAX IV Laboratory, Lund University, Lund 22100, Sweden*
[9]*Department of Physics and Astronomy, University of Bologna, Bologna 40127, Italy*
[10]*PSI Center for Neutron and Muon Sciences CNM, Villigen PSI 5232, Switzerland*

**Tunable quantum materials hold great potential for applications. Of special interest are materials in which small lattice strain induces giant electronic responses. The kagome compounds $AV_3Sb_5$ ($A$ = K, Rb, Cs) provide a testbed for electronic tunable states. In this study, through angle-resolved photoemission spectroscopy, we provide comprehensive spectroscopic measurements of the electronic responses induced by compressive and tensile strains on the charge-density-wave (CDW) and van Hove singularity (VHS) in $CsV_3Sb_5$. We observe a tripling of the CDW gap magnitudes with ~1% strain. Simultaneously, changes of both energy and mass of the VHS are observed. Combined, this reveals an anticorrelation between the unconventional CDW order parameter and the mass of the VHS, and highlight the role of the latter in the superconducting pairing. The substantial electronic responses uncover a rich strain tunability of the versatile kagome system in studying quantum interplays under lattice variations.**

Uniaxial strain possesses a promising potential for engineering the band structure and hence physical properties of quantum materials [1–5]. Near quantum critical points, even small strain stimulus can induce dramatic electronic responses [3–5]. Yet, spectroscopic demonstrations of such effects remain challenging [5].

Recently discovered kagome superconductors $AV_3Sb_5$ ($A$ = K, Rb, Cs) provide an uncharted ground for studies of symmetry-breaking states emerging from electronic correlations [6]. Charge density wave (CDW) [7–9] and superconductivity (SC) [10–12] are currently discussed intensely. The unconventional CDW in $AV_3Sb_5$ is characterized by features such as the chiral responses [13, 14], electronic nematicity [15, 16], and breaking of time-reversal symmetry [12, 17]. Pressure-tunable pairing symmetries [12] and double $T_c$ domes [10] also point to an unconventional nature of SC.

Despite intensive studies, the CDW origin and SC pairing mechanism in $AV_3Sb_5$ are still debated. Both instabilities may be linked to the presence of a high-order van Hove singularity (HO-VHS) [18–20]. That is a saddle point with vanishingly small band curvature. A HO-VHS features power-law divergent density of states (DOS) and further enhanced correlation effects [18]. Its manifestation in the charge-ordered $AV_3Sb_5$ offers a fine quantum tuning knob. Any intimate connections to a HO-VHS would disclose the nature of associated broken-symmetry states. Direct measurements that reveal the interplay of lattice and electronic degrees of freedom are therefore in high demand.

Here, we report angle-resolved photoemission spectroscopy (ARPES) observations of the CDW responses to strain accompanied by a direct tuning of HO-VHS in $CsV_3Sb_5$. Improved ARPES band maps, assisted by density functional theory (DFT), allow a detailed examination of the low-energy electronic structure. Through both uniaxial compressive and tensile strains, we observe a significant response from the CDW order parameter. That is, the energy gap is reduced by a factor of three (from ~150 to 55 meV) as the strain is adjusted from compressive to tensile (±0.6%). Simultaneously, the mass of fermions at a HO-VHS is enhanced, whose energy is driven towards $E_F$. The HO saddle point splits into two upon a compression. Conversely, a nearly ideal HO-VHS is realised with a tensile strain. These observations demonstrate an unconventional strain-sensitive charge modulation with strong electronic anticorrelations coupling to a highly tunable HO-VHS. Given the rising SC $T_c$ with tensile strain, our results of enhanced electronic correlations at $E_F$ suggest the important role of a HO singularity in the SC pairing in CsV$_3$Sb$_5$.

---

* linchun@stanford.edu
† Present address: Stanford Synchrotron Radiation Lightsource, SLAC National Accelerator Laboratory, Menlo Park, California 94025, USA

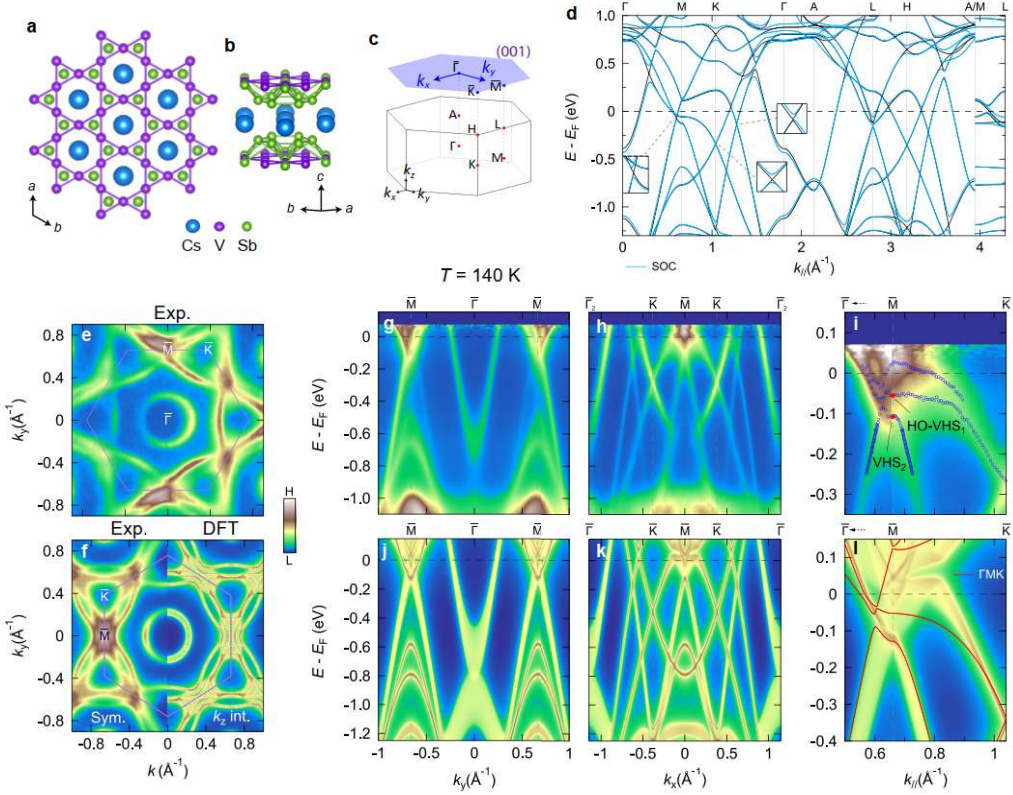

FIG. 1. **Band structure of the kagome superconductor CsV$_3$Sb$_5$ above $T_{CDW}$.** **a,b**, Top and side views of the quasi two-dimensional (2D) crystal structure. **c**, Three-dimensional (3D) Brillouin zone (BZ) and its 2D projection. **d**, Band structure along high-symmetry paths calculated by density functional theory (DFT) without and with spin-orbit coupling (SOC). **e**, Experimental (Exp.) Fermi surface (FS). **f**, Comparison between the experimental and the calculated $k_z$ integrated (int.) FSs. The experimental FS is mirror symmetrized (Sym.) with respect to $\overline{MK}$ and $\overline{\Gamma M}$ which effectively minimizes the matrix-element effects (Supplementary Fig. S2). Purple hexagons in **e,f** indicate the surface BZ. **g,h**, Experimental spectra divided by the Fermi-Dirac function (Supplementary Fig. S3 and Methods) along $\overline{\Gamma M}$ and $\overline{MK\Gamma}$. **i**, Magnified spectrum along $\overline{\Gamma MK}$, where the quasi-particle peaks near $E_F$ are plotted around the $\overline{M}$ point showing two van Hove singularities (VHS$_{1,2}$) as indicated by the red points. VHS$_1$ is of high-order (HO) nature, see details in Methods and Supplementary Fig. S4. **j-l**, Correspondingly calculated $k_z$-integrated dispersions with SOC. Measurement temperature was set at $T = 140$ K, see Methods for more detailed experimental conditions. The colour bar represents ARPES intensities from high (H) to low (L) in arbitrary units.

The quasi two-dimensional (2D) crystal structure of CsV$_3$Sb$_5$ facilitates ARPES observations on (001) surfaces (Fig. 1a-c). Band structure observed above the CDW temperature ($T_{CDW} \sim 94$ K) and first-principle calculations by DFT are presented in Fig. 1d-l. We focus on the spectra near $k_z = 0$ throughout this study, see Supplementary Fig. S1 for the discussions about $k_z$ dispersions. The Fermi surface is composed of a Sb $p$-dominated electron pocket centering the $\overline{\Gamma}$ point and V $d$-dominated kagome-net bands near the Brillouin zone (BZ) boundaries [21] (Fig. 1d-f). In Fig. 1f, we mirror symmetrised the Fermi surface to alleviate photoemission matrix-element effect which alternates spectral weights of different bands when crossing BZ boundaries (Supplementary Fig. S2). As such, the observation matches nicely with $k_z$ integrated DFT (Fig. 1f). Figure 1g,h shows high-symmetry band dispersions which agree well with the DFT (Fig. 1j,k). Note that, all dispersions are divided by a resolution-convolved Fermi-Dirac function to recover single-particle spectral functions allowed by thermal populations (Supplementary Fig. S3). In Fig. 1i,l, we detail the two VHSs near $E_F$. Band structure around the $\overline{M}$ point is extracted from spectral peaks (Supplementary Fig. S4). Polynomial fits for the VHS$_1$ band closer to $E_F$ validate its high-order nature, consistent with previous reports [19, 20]. The main spectral peaks along $\overline{MK}$ disperse across $E_F$, which is identified as a valence band crossing between M and L points in line with the slab calculation in Fig. 1l (see also Fig. 2k). Notably, experiments reveal overall narrower band widths than DFT as indicated by the energy ranges in Fig. 1g-l.

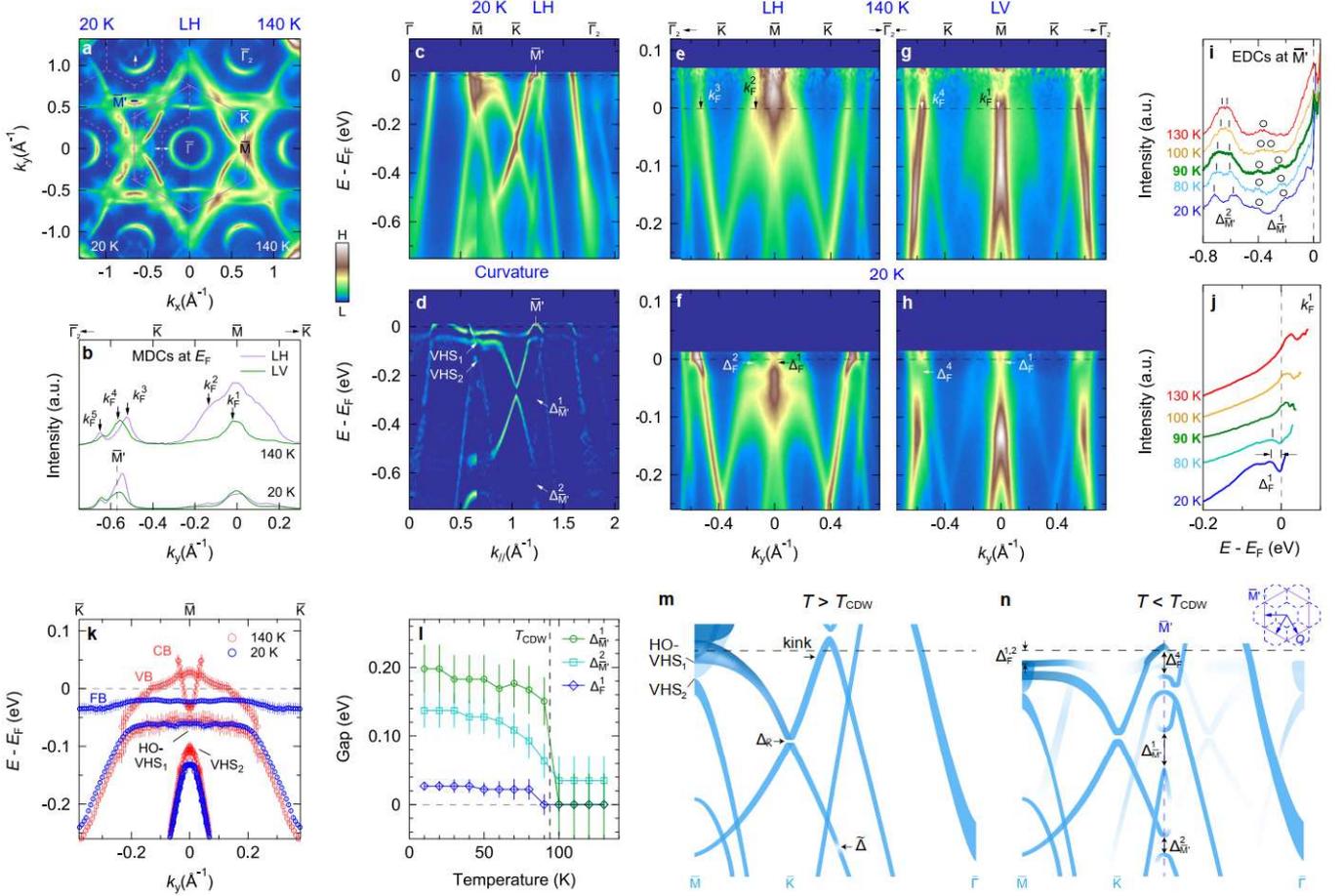

FIG. 2. **Temperature-dependent band structure across the CDW transition.** a, Fermi surfaces acquired at 20 K and 140 K with linear horizontal (LH) light polarization. Solid and dashed hexagons depict the pristine and CDW-reduced surface BZs with $\overline{M}'$ indicating the new BZ boundary. White arrows exemplify CDW band foldings between $\overline{\Gamma}$ and $\overline{M}$ points. b, Polarization-dependent momentum distribution curves (MDCs) at $E_F$ along $\overline{MK\Gamma}$ defining various Fermi momenta. c,d, Band dispersion along $\overline{\Gamma MK\Gamma}$ and its curvature plot (Methods) collected at 20 K with LH polarization. e-h, Temperature- and polarization-dependent low-energy $\overline{MK}$ band structures. i,j, Temperature dependences of energy distribution curves (EDCs) at momentum positions of $\overline{M}'$ and $k_F^1$. The vertical bars or circles in i indicate the peak positions with separations defining the corresponding spectral gaps of $\Delta_{\overline{M}'}^{1,2}$ (Methods). Note that gap $\Delta_F^1$ in j indicated by the arrow only refers to the gap of occupied states below $E_F$ (Methods). k, Temperature-dependent low-energy band structures near VHSs, see details in Methods and Supplementary Fig. S4. Conduction band (CB), valence band (VB), and flat band (FB) are indicated. l, Temperature dependence of energy gaps in i,j. CDW transition temperature ($T_{\text{CDW}} \sim 94$ K) is indicated by the vertical dash line. m,n, Schematics of temperature evolution of $\overline{MK\Gamma}$ band structure deduced from Fig. 1h and c-h. $\widetilde{\Delta}$ indicates a possible fluctuating CDW gap. Band-folding gaps at $\overline{M}'$ are labeled as $\Delta_{\overline{M}'}^i$ ($i$: integer). Fermi momenta and corresponding gaps are marked as $k_F^i$ and $\Delta_F^i$ respectively. All spectra are divided by the Fermi-Dirac function, see Methods for more detailed experimental conditions. The colour bar represents ARPES intensities from high (H) to low (L) in arbitrary units.

In Fig. 2, we present a temperature evolution of band structure across the CDW transition. Side-by-side comparison is shown in Fig. 2a where the Fermi surfaces were collected above and below $T_{\text{CDW}}$. When entering the CDW state, we observe a Fermi-surface reconstruction manifested as the emergence of "shadow" bands (white arrows in Fig. 2a). These folding bands with rotation-symmetry breaking are consistent with the recent observation of electronic nematicity driven by the 3D CDW with $c$-axis $\pi$-phase shift in KV$_3$Sb$_5$ [22].

Low-temperature dispersions along $\overline{\Gamma MK\Gamma}$ is plotted in Fig. 2c,d. The $\overline{M}'$ point indicates the boundary of CDW reconstructed new BZ (dashed hexagons in Fig. 2a). We denote the energy gaps at the $\overline{M}'$ as $\Delta^i_{\overline{M}'}$ (*i*: integer), whose temperature dependences of energy distribution curves (EDCs) and extracted gaps confirm the CDW origin (Fig. 2i,l). A significant fluctuation well above $T_{CDW}$ is suggested by the possible opening of $\Delta^2_{\overline{M}'} \sim 130$ K—consistent with the recent report of a short-range CDW at ~160 K [23]. We note that $\Delta^1_{\overline{M}'}$ tends to close exactly at $T_{CDW}$. Given that DFT calculations indicate the same CDW origin for $\Delta^{1,2}_{\overline{M}'}$ (Supplementary Fig. S5c,d). One possibility is that the weak quasi-particle peaks hinder the detection of $\Delta^1_{\overline{M}'}$ at high temperatures. In Fig. 2e-h, we present the CDW effects on the low-energy states near $E_F$ with Fermi momenta ($k^i_F$, Fig. 2b) and corresponding gaps ($\Delta^i_F$) indicated. With temperature- and light-polarization-dependent spectra in Fig. 2b,e-h, different bands and gaps are clearly resolved. Below $T_{CDW}$, $\Delta^i_F$ opens at $k^i_F$ where $i = 1, 2$, and 4 (Fig. 2f,h,j). The pronounced CDW-folded bands with respect to the $\overline{M}'$ as well as rich energy gaps suggest a long-range charge modulation in the ground state (see also Supplementary Fig. S5a,b). In additon, the Fermi-surface contour is extracted from Fermi momenta in Supplementary Fig. S6. The nesting wave vector of two nearly parallel Fermi-surface sheets in the extended zone scheme is found to coincide with the CDW wave vector reported [8, 9, 19] (see also notes in Supplementary Fig. S6).

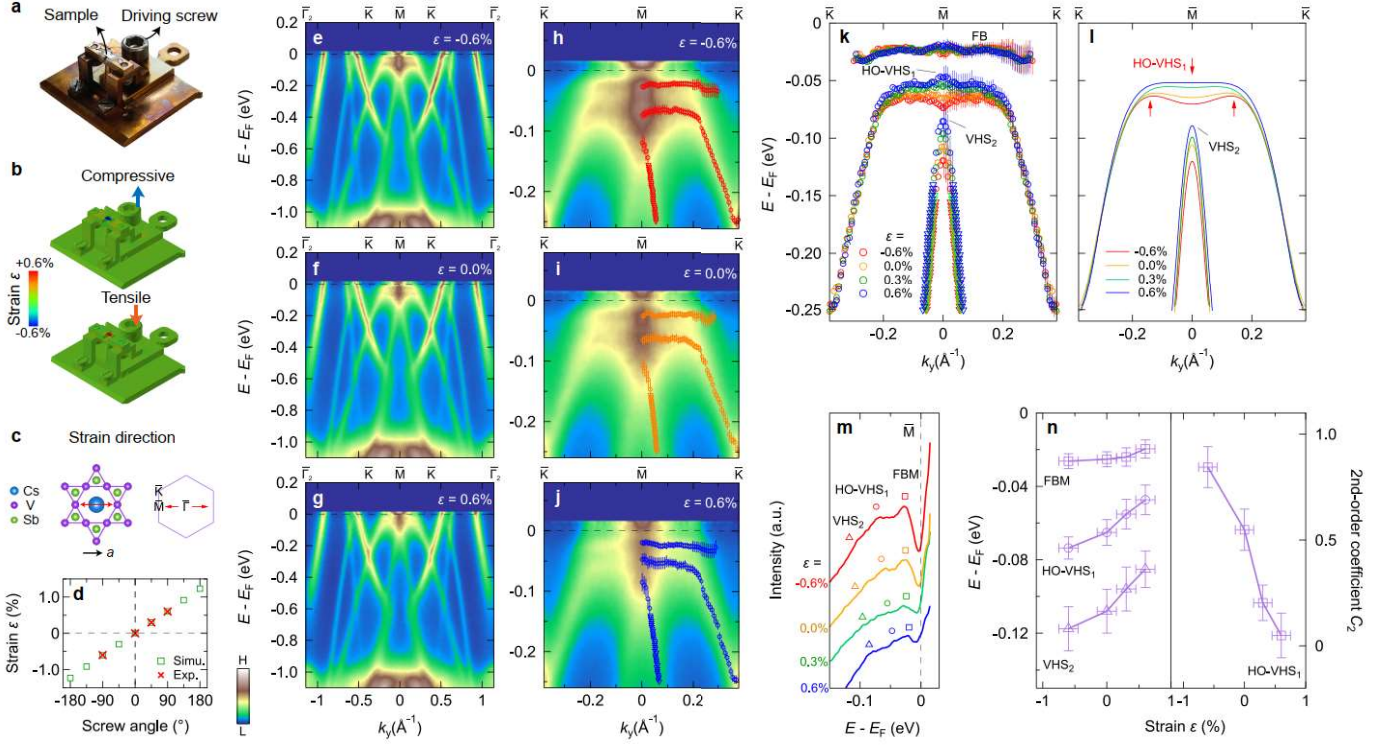

FIG. 3. **Strain tuning of the VHSs.** **a**, *In-situ* strain device designed for compressive and tensile strain ($\epsilon$) applications. **b**, Finite-element simulations for the cases of compressive ($\epsilon < 0$) and tensile ($\epsilon > 0$) strains. **c**, Strain direction in real and reciprocal spaces as indicated by red arrows. **d**, Strain as a function of driving-screw rotation angle estimated from simulation (Simu.). Experimentally accessed *in-situ* strains are marked by the red crosses (Methods). **e-j**, Band dispersions along $\overline{M}\overline{K}\overline{\Gamma}$ collected with *in-situ* nominal strains as indicated. Spectra in **h-j** are magnified around $\overline{M}\overline{K}$ near $E_F$ and symmetrized with respect to $k_y = 0$. Superimposed are the extracted spectral peaks from both EDCs (circle markers) and MDCs (triangle markers), see details in Supplementary Fig. S8. **k**, Band dispersions from the extracted peaks as a function of strain. **l**, Polynomial fits for the two VHS-derived bands in **k** (Supplementary Fig. S9 and Methods). **m**, Strain dependence of the EDCs at $\overline{M}$ point. The peak positions of $VHS_{1,2}$ and flat band at $\overline{M}$ (FBM) are indicated by the respective markers. **n**, Strain-dependent energy positions of $VHS_{1,2}$ and FBM, as well as the second-order fitting coefficient ($C_2$) of $VHS_1$ band that is inversely proportional to its high-order character (Methods). Measurement temperature was set at 10 K and all spectra are divided by the Fermi-Dirac function, see Methods for more detailed experimental conditions. The spectral and strain colour bars represent ARPES intensities from high (H) to low (L) in arbitrary units and longitudinal uniaxial strain from compressive (blue) to tensile (red), respectively.

The gap openings are accompanied by the electronic renormalizations near the VHSs. This is demonstrated by the extracted band structure along $\overline{M}\overline{K}$ (Fig. 2k). A conduction/valence band (CB/VB) and a flat band (FB) are identified at high and low temperatures, respectively. The FB is attributed from the CDW renormalizations around $E_F$ manifested as a continuous gapping of the CB and VB (see also notes in Supplementary Fig. S5).

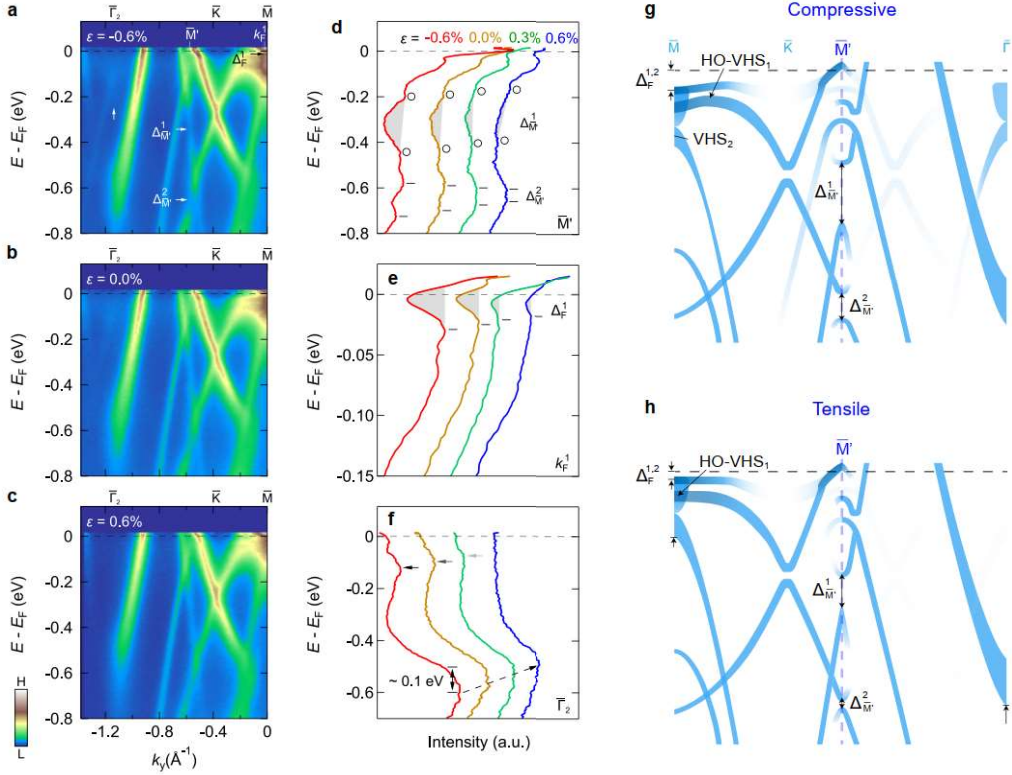

FIG. 4. **Strain tuning of the CDW. a-c**, Magnified $\overline{M}\overline{K}\overline{\Gamma}$ dispersions with strain $\epsilon$ as indicated. CDW-folded bands (vertical arrow) and CDW-related gaps as well as corresponding momentum positions are indicated in **a**. **d-f**, Strain dependences of EDCs at $\overline{M}'$, $k_F^1$, and $\overline{\Gamma}_2$. The circles and bars in **d,e** indicate the positions of quasi-particle peaks or spectral shoulders, whose separations in **d** define the nominal gaps of $\Delta_{\overline{M}'}^{1,2}$. The shaded areas in **d,e** represent spectral-weight suppressions for $\Delta_{\overline{M}'}^1$ and $\Delta_F^1$. See also Methods and Supplementary Fig. S11 for details of EDC-peak and spectral-weight analyses. Horizontal short arrows in **f** mark the CDW folding bands vanishing with tensile strain, while long dashed arrow indicates an energy shift (∼0.1 eV) of the band bottom. **g,h**, Schematics of the band structure in response to uniaxial strain applied along $\overline{\Gamma}\overline{M}$. Vertical arrows indicate either energy gaps or energy shifts. Measurement temperature was set at 10 K and all spectra are divided by the Fermi-Dirac function, see Methods for more detailed experimental conditions. The colour bar represents ARPES intensities from high (H) to low (L) in arbitrary units.

Although both VHSs show a tendency of energy lowering in the CDW state (Fig. 2k). The HO-VHS$_1$ band is abnormally lifted near the $\overline{K}$ point leading to a narrower bandwidth and flattening when approaching the $\overline{M}$ point. This together with the FB indicate further enhanced electronic correlations below $T_{CDW}$ with a mass acquisition. In Fig. 2m,n, we sketch the CDW effects manifested by the band foldings, gap openings, and renormalized heavy fermions near the VHSs.

Next, we turn to observations obtained with strains applied along the $a$-axis. The strain cell, simulations of strain, and strain-calibration curve are shown in Fig. 3a-d. See Methods and Supplementary Fig. S7 for details of strain device and strain evaluation. In Fig. 3e-g, we show the band-structure evolution upon compressive and tensile strain applications. Close-ups along the $\overline{M}\overline{K}$ path near $E_F$ are shown in Fig. 3h-j. Superimposed are the band structures extracted from spectral peaks, which are summarized in Fig. 3k and fitted for the VHS$_{1,2}$-derived bands in Fig. 3l (Supplementary Figs. S8,S9). A tensile strain shifts both VHSs towards $E_F$—see the peaks of EDCs in Fig. 3m. Strikingly, the HO-VHS$_1$ band undergoes a curvature change while maintaining other portions nearly intact. That is, it exhibits a flat and hole-like dispersion near the $\overline{M}$ point upon tensile strain. Whereas a transition to the electron character appears with the compression leading to two new VHSs as indicated by the arrows in Fig. 3l. In Fig. 3n, as a function of strain, we plot the energy positions of VHS$_{1,2}$ as well as the second-order polynomial fitting coefficient ($C_2$) for the VHS$_1$ band. $C_2$ describes the band curvature in the vicinity of VHS$_1$ through $\omega(k)''|_{k=0} = 2C_2$, where $\omega(k)$ is the dispersion relationship. A smaller $C_2$ thus indicates a flatter and higher-order dispersion (Methods or Supplementary Fig. S9). The drastic reduction of $C_2$ suggests that the high-order character of VHS$_1$ is rapidly enhanced with tensile strain (Fig. 3n, see reproducibility in Supplementary Fig. S10).

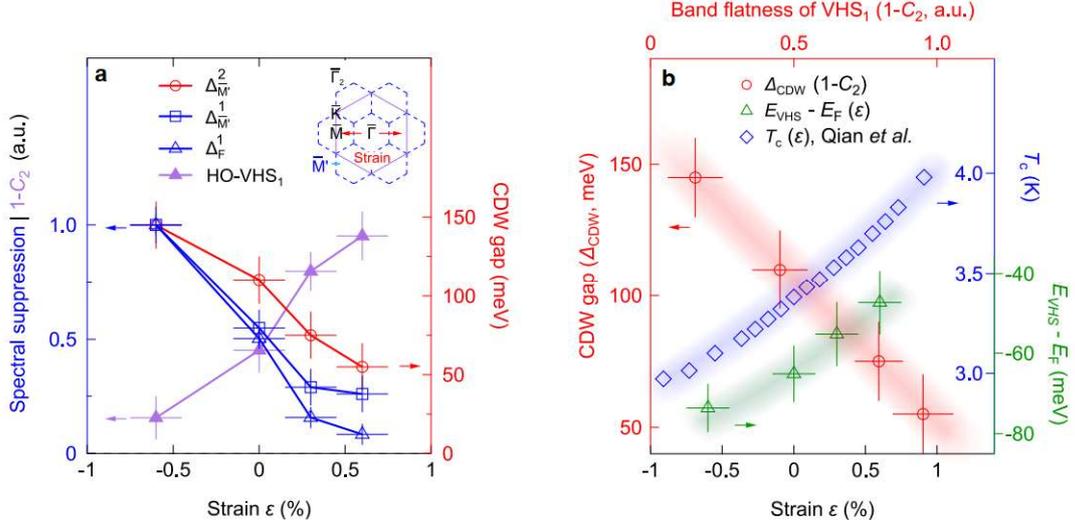

FIG. 5. **Strain response of CDW order parameter and its anticorrelation with HO-VHS. a**, Energy gaps and normalized spectral-weight suppressions as a function of uniaxial strain $\epsilon$. Plotted together is the HO character of VHS$_1$ represented by $1-C_2$, demonstrating its band flatness or qualitatively the normalized effective mass (purple markers). The inset shows a schematic CDW BZ and the direction of applied strain as indicated by the arrow. **b**, Strain-dependent CDW gap ($\Delta_{\text{CDW}}$) versus the band flatness of VHS$_1$ ($1-C_2$). The energy positions of HO-VHS$_1$ ($E_{\text{VHS}} - E_F$, Fig. 3n) and SC $T_c$ (Qian et al. [24]) are plotted as a function of strain for comparisons. Coloured shadings are guides for the eyes. See Methods for the error bars.

Strain effects on the CDW-related electronic states along $\overline{\text{MK}}\overline{\Gamma}$ are shown in Fig. 4a-c. Followed in Fig. 4d-f are strain-dependent EDCs at $\overline{\text{M}}'$, $k_F^1$, and $\overline{\Gamma}_2$, with EDC-peak and spectral-weight analyses (Methods and Supplementary Fig. S11). A direct CDW response to strain is observed—both the band-folding gap at $\overline{\text{M}}'$ ($\Delta_{\overline{\text{M}}'}^{1,2}$) and Fermi-surface gap at $k_F^1$ ($\Delta_F^1$) fade away as strain changes from compressive to tensile. Meanwhile, the analysis for the peaks below $E_F$ at $k_F^1$ (bars in Fig. 4e) shows a reducing coherence (Supplementary Fig. S11c). This observation is in line with the CDW suppression upon tensile straining since the peak is associated with the CDW-renormalized FB (Fig. 2k and Supplementary Fig. S11d-f). Such a CDW suppression is further supported by a simultaneous diminishing of the CDW-folded bands near $\overline{\Gamma}_2$ (horizontal arrows in Fig. 4f, see reversibility in Supplementary Fig. S10). Additionally, the band bottom at $\overline{\Gamma}_2$ is lifted by ~0.1 eV (dashed arrow in Fig. 4f). In Fig. 4g,h, we summarise the overall band-structure evolution under uniaxial strains, featured by the tuning of VHSs (Fig. 3) and a pronounced CDW response.

To quantify the changes of CDW order parameter, Fig. 5a shows strain evolutions of gap magnitudes—defined as either quasi-particle peak separation ($\Delta_{\overline{\text{M}}'}^2$) or spectral-weight suppression ($\Delta_{\overline{\text{M}}'}^1$ and $k_F^1$ as represented by the grey shaded areas in Fig. 4d,e, see Methods). Regardless of definitions, the CDW gaps display a substantial strain response, i.e., the gap magnitudes change by a factor of three with only ~1% strain variation.

The upward shifts of VHSs at the $\overline{\text{M}}$ point agree well with the in-plane strain calculations with a constant c-axis [25]. By contrast, the large energy lifts of Sb-$p_z$ pocket at the $\overline{\Gamma}$ point (Fig. 4f,h) are consistent with the hydrostatic-pressure calculations dominated by the fast shrinkage of $c$ lattice parameter [26]—reminiscent of the charge transfer of $p_z$ orbitals in IrTe$_2$ due to strain-modified interlayer couplings [27]. These results indicate that both in-plane and out-of-plane lattice variations play important roles in pressure tuning the band structure. Furthermore, the distinct energy changes for different bands rule out a simple chemical-potential shift and suggest an orbital-selective bandwidth transfer. It is shown that the $T_{\text{CDW}}$ can be strain-tuned [24]. We plot the mean-field CDW gap deduced by $\Delta(\epsilon) = 3.5 k_B T_{\text{CDW}}(\epsilon)$ in Supplementary Fig. S12. Our results support a much faster CDW-gap suppression towards tensile direction, manifested by not only a triple reduction but also the absolute change of ~95 meV (from ~150 to 55 meV for $\Delta_{\overline{\text{M}}'}^2$) with only 1% strain (see also Fig. 5a). A substantial strain response of the CDW order parameters is thus spectroscopically visualised.

The observed CDW response to strain perturbations points to the unconventional nature of charge modulations in CsV$_3$Sb$_5$. Lattice distortions with a star-of-David pattern or its inverse counterpart are reported to coexist [28, 29] (see also the notes in Supplementary Fig. S5). CDW orders with different c-axis stackings are also observed [30, 31]. Such an intertwist of the CDW supercells with significant interlayer couplings [27, 28] is likely sensitive to small lattice variations, in particular for a competing relationship [31]. These studies thus imply a possible strain-induced change of the charge-modulation patterns or interlayer stacking phases leading to the CDW response observed here. It is worth noting that, the strain-dependent observations are obtained at ~10 K. The band structure exhibits nematic band foldings (Fig. 2a) [22]. Therefore, the reported CDW-driven electronic nematicity [15] or unidirectional charge order [16, 32] may also contribute to the observed gap openings and react with the uniaxial strain. The observations of nested Fermi-surface gaps and VHSs favour an electronic origin of the CDW (Fig. 2k and Supplementary Fig. S6). This further conforms the recent Ginzburg-Landau minimal model proposing a charge bond order (CBO) driven by the sublattice-assisted Fermi-surface instability [33]. Threefold CBO has a tendency of unfolding into various orders. A chiral charge order is energetically preferred. The lattice perturbation can be a tuning knob for such phase transitions. Our results thus provide a natural interpretation for the conflicting reports of chiral CDW states [13, 34], which likely stem from the residual local strain. Additionally, our current studies also suggest the indispensable role of electron-phonon coupling (EPC) bridging the lattice perturbations to electronic responses. Significant EPC kinks were indeed found in $AV_3Sb_5$ [35, 36] (Supplementary Fig. S5b). We leave its detailed strain effects for a future work.

A minor lattice perturbation is generally not expected to generate giant changes for a system stabilizing at its minimum free-energy state. For significant changes to happen, the perturbation must couple with the electronic states which requires suitable matrix elements. In $CsV_3Sb_5$, CDW strongly distorts the lattice by up to 1% [37]. In turn, lattice variation couples to the CDW gap giving rise to the strain-sensitive CDW order parameter. Consequently, the gaps initially opened by the CDW (see our DFT calculations in Supplementary Fig. S5), experience a rapid enhancement when strain is applied. Here, we highlight the important role of proximity to a high-order singularity [18]. The strain response of HO-VHS$_1$ is manifested by its rapid mass enhancement (band flattening in Fig. 3l) with a drop of the band curvature ($C_2$ in Fig. 3n). Qualitatively, the band flatness or normalized effective mass can be represented as $1-C_2$. As such, we plot $1-C_2$ as functions of strain (Fig. 5a) and the CDW gap (Fig. 5b). A dramatic enhancement of the mass is accompanied simultaneously with an instant CDW suppression towards the tensile direction (Fig. 5a). The mass enhancement of HO-VHS$_1$ reveals a nearly ideal HO-VHS achieved by a fine strain tuning (see also Fig. 3l). The anticorrelation between the HO-VHS$_1$ and CDW order parameter (Fig. 5b) naturally explains the splitting of VHS$_1$ arising from CDW renormalizations [38] (Fig. 3l). These observations demonstrate a rare example of strain sensitive HO-VHS with strong electronic correlations coupling to an unconventional charge modulation.

The CDW and SC orders are believed to compete in the $CsV_3Sb_5$ system [10, 24, 28]. The anticorrelation between the CDW and HO-VHS$_1$, therefore, implies a supportive connection of the latter to the underlying SC. Indeed, the SC $T_c$ is reported to increase under tensile strain (Fig. 5b) [24]. Correspondingly, we uncover an enhanced DOS and electronic correlations at $E_F$, arising from the mass enhancement and energy lift of the HO-VHS$_1$ (Fig. 5b). The strain-$T_c$ study [24] points to a prominent role of $c$-lattice constant in line with hydrostatic-pressure effects (see also notes in Supplementary Fig. S12). Here, our observations of kagome-driven VHSs and Sb-$p$ band indicate that both in- and out-of-plane lattice variations are important in strain tunings. More importantly, these observations suggest a favourable connection between the HO-VHS$_1$ and the SC order. Under electron-electron SC mechanism, it is known that a high/diverging DOS leads to a propensity for instabilities [39]. Our results thus shed light on the important role of a HO-VHS and electron-electron interactions in the SC pairing mechanism of $CsV_3Sb_5$.

As a reversible, flexible, disorder-free, and single-parameter tuning knob, uniaxial strain has demonstrated its highly desirable role in engineering quantum materials [1–5]. Given the numerous reports showing a strain sensitivity in $AV_3Sb_5$ [14, 40, 41], our direct observations of electronic responses reveal a rich strain tunability on such a versatile platform inspiring further studies of quantum interplays under lattice perturbations.

# Methods

**Sample.** High-quality $CsV_3Sb_5$ single crystals were grown by the modified self-flux method [6, 32]. To facilitate uniaxial strain application, the crystals were cut into a rectangular shape with in-plane dimension ~1 × 0.8 mm$^2$.

**ARPES.** ARPES measurements were performed at the beamline I05 of Diamond Light Source in UK and beamline Bloch of MAX IV Laboratory in Sweden. Samples were mounted using silver epoxy and cleaved by standard top-post method *in-situ*. Beam spot sizes were set at ~50 × 50 and 10 × 5 μm$^2$, respectively. Photon energies were scanned from ~40 eV to 100 eV to assign the high-symmetry $k_z$ points (Supplementary Fig. S1). To access dominant spectral weights at $k_z \sim 0$ $\pi/c$, photon energy was set at 82 eV with overall energy resolution 12–16 meV. Both linear horizontal and vertical light polarizations were utilised to obtain the entire band structure with high contrasts (Fig. 2 and Supplementary Fig. S4). Electron analysers with vertical slits are MBS A-1 from MB Scientific AB and DA30L from ScientaOmicron, respectively, with angular resolutions 0.1–0.2°. Photoelectron emission angles were scanned by a step of 0.1° to precisely locate the momentum positions. Low (high) temperature measurements were conducted in vacuum conditions with pressures better than $1.5 \times 10^{-10}$ ($2 \times 10^{-10}$) mbar and $8 \times 10^{-11}$ ($1 \times 10^{-10}$) Torr, respectively. Strain measurements are conducted at the base temperature, see also Supplementary Table S1 for experimental conditions.

**Analysis.** Figure 1i,l and Fig. 2c,d are composed by two spectra collected along orthogonal paths. Owing to the surface sensitivity and quasi-2D structure, surface BZ is utilised (Fig. 1c). *Spectral symmetrization* is adopted in Fig. 1e,g and Fig. 2k to minimize matrix-element effect in Fermi surfaces, and in Fig. 3h-j to reduce noise. Fermi-surface symmetrizations were done by assuming only mirror symmetries with respect to $\overline{MK}$ and $\overline{\Gamma M}$ owing to the possible nematic states at low temperature. *Fermi-Dirac function division* (FDD) is utilised throughout this study, i.e., the spectra are divided by a Fermi-Dirac function that is convolved with a Gaussian representing the energy resolution [42], which effectively removes the Fermi cutoff and recovers the single-particle spectral function to the extent allowed by thermal population (Supplementary Fig. S3). On the other hand, energy gaps revealed after recovering spectral functions do not assume particle-hole symmetry which is typically broken in the CDW state. *Curvature or second-derivative analysis* [43] is adopted to enhance band visualisation in Fig. 2d. These methods inevitably produce "peaks" or "flat bands" below $E_F$ due to Fermi cutoffs (Supplementary Fig. S3e,g). To alleviate such artifacts near $E_F$, FDD spectra are utilised when performing the analysis. *Spectral peaks* of energy/momentum distribution curves (EDCs/MDCs) are extracted by analysing either first/second derivatives [44] or fittings to the multi-peak Lorentzian statistics (Supplementary Fig. S11), with error bars estimated in accordance with the smoothnesses and widths of peaks (Supplementary Fig. S4,S8). *CDW gap magnitudes* of $\Delta_{\overline{M}'}^1$ and $\Delta_F^1$ are defined as spectral-weight suppressions, represented by the grey shaded areas in Fig. 4d,e. They are defined by the enclosed areas of EDCs and line segments either between spectral peaks in Fig. 4d or of peak-height extrapolations in Fig. 4e. ARPES spectral weights contain rich information of intertwined states in correlated systems [42, 45, 46], which is also in general proportional to DOS at a given condition. A spectral-weight suppression is therefore directly linked to DOS depletion representing a phase coherence or order parameter that can be utilised to quantify energy gaps especially in a partial gap scenario [45]. In addition, because a CDW gap at $E_F$ is not necessarily particle-hole symmetric [47], $\Delta_F^1$ in Fig. 2j only refers to the gap of occupied states. The uncertainty of the absolute gap size at $E_F$ is therefore another reason to utilize spectral-weight suppression when demonstrating the strain effect on the Fermi surface (see also notes in Supplementary Fig. S11). *Fermi momenta* $k_F$ are obtained from Lorentzian fittings to MDCs at $E_F$ at high temperatures. $k_F 1,2$ in Fig. 2e,g,b are determined in conjunction with the CB and VB $E_F$ crossings in Fig. 2k. *Polynomial fits* of VHSs were done using $\omega(k) = C_0 + C_2 \cdot k^2 + C_4 \cdot k^4 + ...$ with $C_n$ denoting the $n$-th order fitting coefficients (Supplementary Fig. S9). The band flatness near the vicinity of VHS is controlled by $\omega(k)''|_{k=0} = 2C_2$. A flatter VHS is therefore associated with smaller $C_2$ and non-zero high-order terms, i.e., HO-VHS. The error bars of $C_2$ in Fig. 5a are fitting standard deviations.

**Strain device.** Our new uniaxial strain device is based on a customized bendable sample platform that enables *in-situ* applications of more than ± 3% compressive or tensile strain. The design is composed of a T-shape platform, a step-like actuator, a driving screw with a circlip, and a customized omicron holder. All components are non-magnetic and typically made of BeCu with excellent elastic, thermal, and electric properties. As shown in Fig. 3a,b, by tightening the driving screw, the actuator is pushed down and therefore enforces a downward bending of the edge-connected platform providing tensile strain along the platform. Loosing the driving screw lifts the actuator driven by the underneath circlip which upward bends the platform realizing a compression. Such a driving mechanism enables a compact and flexible cell that fits spectroscopic manipulators.

**Strain evaluation.** The *in-situ* strain controls were realised by rotating the driving screws using equipped wobble stick. Strain evaluation follows the previous report, in which a precise connection between the real strain magnitude measured by the commercial gauges and the finite-element simulated strain size was well established (Supplementary Fig. S7e) [5]. Therefore, the strain magnitudes at the sample position as a function of driving-screw angle can be simulated as shown in Fig. 3d and Supplementary Fig. S7a-d. The strain-angle curve is simulated at the device center, where the strain gradient is small and homogeneous ARPES data are collected. The strain magnitudes *in-situ* controlled in ARPES measurements were then evaluated by interpolating the strain–angle curve at the corresponding screw angles (red markers in Fig. 3d). The error bars for the strain magnitude in Fig. 3n and Fig. 5a include the uncertainties of screw angles and thermal contractions.

**DFT calculation.** Density functional theory calculations have been performed using the Vienna ab-initio simulation package (VASP) [48], within the projector-augmented plane-wave (PAW) method [49]. When dealing with the $CsV_3Sb_5$ unit cell, the Kohn-Sham wave functions are expanded into plane waves up to an energy cutoff of 600 eV, using the PBE-GGA functional to handle the exchange-correlation potential [50]. The Brillouin zone has been sampled with a 12×12×8 Γ-centered $k$-mesh including spin-orbit coupling self-consistently, when present. The relaxation of the electronic degrees of freedom was considered converged when the energy difference among two steps was equal or smaller than $1.0^{-6}$ eV. To compute the partial occupancies it was used a Gaussian smearing with a width of 0.02 eV. When considering the 2×2×1 star of David and tri-hexagonal supercells, the Kohn-Sham wave functions are expanded into plane waves up to an energy cutoff of 400 eV. Band structures have been visualized using the VASPKIT postprocessing tool [51]. VESTA [52] has been used to visualize crystal structures.

## Data availability

All data that support the findings of this study are included in the main text and supplementary information. Other data are available from the corresponding author upon request.

---

## Acknowledgements

C.L., J.K., and J.C. acknowledge support from the Swiss National Science Foundation (Projects No. 200021_188564). C.L. and M.M.D. thank the support from Forschungskredit of the University of Zurich under grant numbers FK-22-119 and FK-22-085, respectively. A.C. acknowledges support from PNRR MUR project PE0000023-NQSTI. A.C. and G.S. acknowledge the Gauss Centre for Supercomputing e.V. (https://www.gauss-centre.eu) for funding this project by providing computing time on the GCS Supercomputer SuperMUC-NG at Leibniz Supercomputing Centre (https://www.lrz.de). A.C. and G.S. are also grateful for funding support from the Deutsche Forschungsgemeinschaft (DFG, German Research Foundation) under Germany's Excellence Strategy through the Würzburg-Dresden Cluster of Excellence on Complexity and Topology in Quantum Matter ct.qmat (EXC 2147, Project ID 390858490) as well as through the Collaborative Research Center SFB 1170 ToCoTronics (Project ID 258499086). O.K.F is supported by the Swedish Research Council (VR) via a Grant 2022-06217 and the Foundation Blanceflor 2023 fellow scholarship. J.K. is supported by a Ph.D. fellowship from the German Academic Scholarship Foundation. H.L. is supported by National Key R&D Program of China (Grants Nos. 2022YFA1403800 and 2023YFA1406500), National Natural Science Foundation of China (Grants No. 12274459), Beijing Natural Science Foundation (Grant No. Z200005). Z.G. acknowledges support from the Swiss National Science Foundation (SNSF) through SNSF Starting Grant (No. TMSGI2_211750). R.T., G.S. and T.N. acknowledge support from Deutsche Forschungsgemeinschaft (DFG, German Research Foundation) and the Swiss National Science Foundation (Project 200021E_198011), respectively, as part of the FOR 5249 (QUAST). We acknowledge MAX IV Laboratory for time on Beamline Bloch under Proposal 20221475. Research conducted at MAX IV, a Swedish national user facility, is supported by the Swedish Research council under contract 2018-07152, the Swedish Governmental Agency for Innovation Systems under contract 2018-04969, and Formas under contract 2019-02496. We acknowledge Diamond Light Source for time on Beamline I05 under Proposals SI30650 and SI33528.

## Author contributions

J.C. and C.L. conceived the project. C.L. carried out the ARPES experiments with supports from O.K.F., J.K., J.C., D.C., M.L., C.P., B.T., A.L., M.W., T.K., and C.C.. C.L. designed the uniaxial strain device with supports from J.C.. C.L. analysed the data and performed the strain evaluations. A.C. calculated the band structure with supports from M.M.D., D.D.S., R.T., G.S., and T.N.. H.L. and Z.G. provided the single crystals. C.L., A.C., and J.C. wrote the manuscript with inputs from all authors. All authors discussed the results and commented on the manuscript.

## Competing interests

The authors declare no competing interests.